\documentclass{jetpl}

\twocolumn
\lat

\usepackage{graphicx}

\def \be#1\ee {\begin{equation}#1\end{equation}}
\def \bem#1\eem {\begin{multline}#1\end{multline}}
\def \beg#1\eeg {\begin{gather}#1\end{gather}}
\def \bea#1\eea {\begin{align}#1\end{align}}

\newcommand{\eps}{\varepsilon}
\newcommand{\vphi}{\varphi}
\newcommand{\tQ}{\tilde Q}

\newcommand{\tEg}{\tilde E_g}
\newcommand{\ETh}{E_{\mathrm{Th}}}

\newcommand{\calF}{\mathcal{F}}
\newcommand{\tx}{\hat\tau_1}
\newcommand{\ty}{\hat\tau_2}
\newcommand{\tz}{\hat\tau_3}

\DeclareMathOperator{\Tr}{Tr}
\DeclareMathOperator{\tr}{tr}

\title{Coulomb Effects in Nanoscale SINIS Junction}
\rtitle{Coulomb Effects in Nanoscale SINIS Junction}
\sodtitle{Coulomb Effects in Nanoscale SINIS Junction}

\author{P.\,M.\,Ostrovsky\thanks{e-mail: ostrov@itp.ac.ru} and M.\,V.\,Feigel'man}
\rauthor{P.\,M.\,Ostrovsky, M.\,V.\,Feigel'man}
\sodauthor{Ostrovsky, Feigel'man}

\address{L. D. Landau Institute for Theoretical Physics, Kosygina 2, Moscow, 119334, Russia}

\abstract{We study the system of two superconductors connected by a small
normal grain. We consider the modification of the Josephson effect by the
Coulomb interaction on the grain. Coherent charge transport through the
junction is suppressed by Coulomb repulsion. An optional gate electrode may relax the
charge blocking and enhance the current  leading to the single
Cooper pair transistor effect. Temperature dependences of critical current and of
the minigap induced in the normal grain by the proximity to superconductor are studied.
Both temperature and Coulomb interaction  suppress critical current and minigap but their
interplay may lead to the nonmonotonous and even reentrant temperature
dependence.}


\begin{document}

\maketitle

Nanoscale SINIS junction consists of a small normal metallic grain connected to
two superconductive leads by tunnel junctions. The Josephson effect in this
system is provided by the Andreev reflection processes at the contacts: Cooper
pairs enter the normal part of the junction and propagate as Cooperons. This
allows the transport of Cooper pairs from one lead to another establishing a
supercurrent. Another manifestation of the Andreev mechanism is the appearance
of a minigap in the spectrum of the normal grain: the proximity effect. This
minigap is a result of a non-zero Cooper pairs density come from
superconductors. The mentioned effects rely on the phase conservation in the
grain. Charging effects lead to fluctuations of the phase and break the
coherent electron transport as well as the induced minigap. The interplay
between proximity and charging effects in a normal grain connected to one
superconductor was recently studied in~\cite{OSF04}. Here we apply the same
formalism to the system with two leads and consider the Josephson effect. We
also study temperature dependence of the minigap and Josephson current, and
find some unexpected reentrant behavior in a certain range of parameters.

We consider  tunnel junctions between normal  grain and left (L) and right (R)
superconductors characterized by large [in units of $e^2/\hbar$] normal-state
conductances $G_{L,R} \gg 1$ and (geometric) capacitances $C_{L,R}$. The gate
electrode is coupled to the grain by the capacitance $C_g$. Mean level spacing
in the grain $\delta$ is the smallest energy scale of the system while the
superconductive gap $\Delta$ in the leads and Thouless energy $\ETh$ of the
grain are largest ones. We assume that Andreev conductances of both contacts
are small, $G^A_{L,R} \leq 1$ (together with conditions $G_{L,R} \gg 1$ it
means that our junctions contain many weakly-transparent channels, quantitatve
estimates will be provided below). The proximity and charging effects in the
grain are characterized by the bare minigap width $E_{g0} = (G_L +
G_R)\delta/4$ and Coulomb energy $E_C = e^2/(2C)$ with $C = C_L + C_R + C_g +
\Delta C$ being the total capacitance of the grain. Here $\Delta C =
\frac{e^2}{2\Delta}(G_L + G_R)$ is the contribution to capacitance coming from
virtual quasiparticle tunnelling~\cite{LarkinOvchinnikov83}. We assume that
$\delta \ll (E_{g0}, E_C) \ll (\Delta, \ETh)$. With the help of the dynamical
(in imaginary time) sigma-model in replica space~\cite{Finkelstein90}, and the
adiabatic approximation for charging effects developed in~\cite{OSF04}, we
study the current-phase relation of SINIS junction as well as the dependence of
the critical current on temperature and gate voltage $V_g$.

Electronic properties of the normal grain are characterized by its Green
function. To capture  proximity induced correlations one uses the matrix Green
function in Nambu-Gor'kov representation. The sigma-model operates with the
matrix field $Q$ which apart from Nambu-Gor'kov structure carries two Matsubara
energies and two replica indices. The standard Green function can be extracted
from the diagonal in energies element of matrix $Q$ by replica averaging.
Charging effects are described by the fluctuating scalar field $\phi$
corresponding to the electric potential and also carrying the replica index in
the sigma-model formalism. The action for the two variables $Q$ and $\phi$
reads
\bem
 S[Q,\phi]
  = -\frac{\pi}{\delta}\Tr \left[
      (\eps\tz+\phi)Q
    \right]\\
    -\frac{\pi}4 \Tr[(G_LQ_L + G_RQ_R)Q]
    + \sum_a\int\limits_0^{1/T} d\tau\frac{(\phi^a - eV_g)^2}{4E_C}.
 \label{actionQphi}
\eem
Here $\hat\tau_i$ are the Pauli matrices in Nambu-Gor'kov space. The trace
operation implies summation over all possible variables including replica
indices and integration over energies. The equilibrium superconductive matrices
$Q_{L,R}$ for the leads are diagonal in both energies and replicas and have the
form $Q_{L,R}(\eps) = \tx\cos\vphi_{L,R} + \ty\sin\vphi_{L,R}$ in Nambu-Gor'kov
space with $\vphi_{L(R)}$ being the superconductive phase of the left (right)
lead. This expression is valid at energies well below $\Delta$. The
contribution from higher energies~\cite{LarkinOvchinnikov83} has already been
taken into account by renormalization of the capacitance: $C \mapsto C +
e^2(G_L+G_R)/(2\Delta)$.

To exclude fast fluctuations due to shifts of the electron band by the
potential $\phi$ from the matrix $Q$ we perform the following change of
variables~\cite{KamenevAndreev99}
\be
 \phi^a(\tau)
  = \dot K^a(\tau),
 \quad
 Q^{ab}_{\tau\tau'}
  = e^{i\tz K^a(\tau)} \tQ^{ab}_{\tau\tau'} e^{-i\tz K^b(\tau')}.
\ee
The phase $K$ is determined up to a constant, which will be fixed later to
simplify further analysis. With new variables the action takes the form
\bem
 S[\tQ,K]
  = -\frac{\pi}{\delta}\Tr(\eps\tz\tQ)
    +\int_0^{1/T} d\tau \biggl[
      \frac{(\dot K - N)^2}{4E_C}\\
      -\frac{2\pi E_g(\vphi)}{\delta}\left(
        \tQ^{(1)}_{\tau\tau}\cos 2K
        +\tQ^{(2)}_{\tau\tau}\sin 2K
      \biggr)
    \right]
 \label{action}
\eem
In this formula we put $\vphi = \vphi_L - \vphi_R$, denote $\tQ^{(i)} =
\tr(\hat\tau_i\tQ)/2$, $N = C_gV_g/e$ and introduce bare phase-dependent
minigap
\be
 E_g(\vphi)
  = \frac{\delta}4 \sqrt{G_L^2 + G_R^2 + 2G_LG_R\cos\vphi}.
\label{Egphi}
\ee

The expression~(\ref{action}) is very similar to the action for an SIN system
with one superconductive lead~\cite{OSF04}. The only difference is the phase
dependence of $E_g$. Further calculation will essentially follow the procedure
of Ref.~\cite{OSF04}. The key idea is \emph{the adiabatic approximation} based
on the separation of characteristic frequencies of matrix $\tQ$ and phase $K$
ensured by the inequality $E_C \gg \delta$. Characteristic timescale of the
variable $K$ fluctuations is much shorter than that of electronic degrees of
freedom, thus we  integrate the action over $K(t)$ regarding $\tQ$ as a
time-independent matrix (it depends only on the difference of its two time
arguments). Than we apply \emph{the saddle point approximation} to the
$K$-averaged action. The justification of this approximation will be provided
below.

We parametrize the time-independent matrix $\tQ$ by an angle $\alpha$:
\be
 \tQ(\eps)
  = \tz\cos\alpha(\eps) + \tx\sin\alpha(\eps).
\ee
The $\ty$-term here is eliminated by the proper choice of the constant in the
definition of $K$. Inserting this expression into~(\ref{action}) we derive the
Hamiltonian controlling the dynamics of the phase $K$:
\be
 \hat H
  = E_C\left[(-i\partial/\partial K - N)^2 - 2q(\vphi)\cos2K \right].
 \label{hamilton}
\ee
All physical quantities depend periodically on $N$. It is convenient to assume
that $|N| < 1/2$. The parameter $q(\vphi)$ is expressed in terms of the angle
$\alpha(\eps)$
\be
 q(\vphi)
  = \frac{\pi E_g(\vphi)T}{E_C\delta}
    \sum_{\eps_n} \sin\alpha(\eps_n).
 \label{q}
\ee
This sum diverges logarithmically and should be cut off at $|\eps| \sim
\Delta$. For large values of $q$ the phase $K$ is nearly localized in the
minima of cosine potential and the fluctuations are weak. In the opposite case
fluctuations of the phase get strong and proximity effect is mostly suppressed.
Thus the parameter $q$ quantifies the strength of proximity coupling competing
with the charging effect.

With the derived Hamiltonian we are able to calculate the free energy of the
$K$ degree of freedom $F(q,T)$ and then extract the total free energy of the
system from the action~(\ref{action})
\be
 \calF
  = -\frac{2\pi T}{\delta} \sum_{\eps_n} \eps_n\cos\alpha(\eps_n) + F(q,T).
 \label{freeen}
\ee
The equilibrium value of $\alpha(\eps)$ is determined by the minimum of this
free energy functional (saddle-point approximation). This gives
$\tan\alpha(\eps) = \tEg/\eps$ with $\tEg$ obeying the self-consistency
equation
\be
 \frac{\tEg}{E_g(\vphi)}
  = -\frac1{2E_C}\frac{\partial F}{\partial q}.
 \label{selfconsist}
\ee
This $\tEg$ is the minigap appearing in the spectrum of the normal grain. The
estimation of matrix $tQ$ fluctuations justifies the saddle-point approximation
provided $\tEg \gg \delta$. The system of two equations~(\ref{q})
and~(\ref{selfconsist}) determines two parameters, $q$ and $\tEg$. A trivial
solution $q = \tEg = 0$ always exists. It is analogous to the normal state
which is the stationary point (local maximum) of the superconductor free
energy. We are looking for a non-trivial solution leading to nonzero value of
the minigap $\tEg$. Once this solution exists it has lower energy than the
trivial solution.

After solving the equations we can calculate the free energy~(\ref{freeen}) and
all physical properties of the junction. We are interested in the current-phase
relation given by the identity $I(\vphi) = (2e/\hbar)(d\calF/d\vphi)$. Using
the self-consistency relation~(\ref{selfconsist}) and the identity~(\ref{q}) we
express the current as
\be
 I(\vphi)
  = \frac{e\delta^2E_C\tEg q}{4\hbar E_g^3(\vphi)}G_LG_R\sin\vphi.
 \label{current}
\ee

Below we consider analytically two limiting cases of weak ($q \gg 1$) and
strong ($q \ll 1$) charging effect and then discuss the numerical results for
arbitrary $q$. The spectrum of the Hamiltonian~(\ref{hamilton}) is given by the
characteristic values of the Mathieu equation which is elementary solved in
these two limits. We first calculate the current at zero temperature taking the
ground state of~(\ref{hamilton}) for the free energy $F$.

\emph{Weak Coulomb blockade.} When charging effects are weak and the parameter
$q$ is large the phase $K$ is localized near $0$ or $\pi$ in the minima of the
cosine potential. The applied gate voltage is ineffective in this case.
Expanding the potential to the second order near its minimum we find the ground
state energy $E_0 = E_C(-2q + 2\sqrt q)$. The pair of
equations~(\ref{q},\ref{selfconsist}) can be solved iteratively. In the
considered regime the minigap is slightly suppressed in comparison with its
bare value $E_g(\vphi)$. First, we estimate $q$ substituting $E_g(\vphi)$ in
the r.h.s. of~(\ref{q})
\be
 q_0
  = \frac{E_g^2(\vphi)}{E_C\delta} \log(\Delta/E_{g0}).
\ee
Here we neglect the $\vphi$-dependence of $E_g$ in the argument of logarithm.
At the next iteration we put $q_0$ in the r.h.s. of~(\ref{selfconsist}) and
then refine the value of $q$ inserting the calculated $\tEg$ into~(\ref{q}):
\be
 \tEg
  = E_g(\vphi)(1-1/2\sqrt{q_0}),
 \quad
 q
  = q_0(1-1/2\sqrt{q_0}).
\ee
With this $q$ and $\tEg$ we calculate the Coulomb correction to the current
using~(\ref{current})
\beg
 I(\vphi)
  = I_0(\vphi)\left[1 - \frac1{\sqrt{q_0}}\right],
 \label{Iweak}
 \\
 I_0(\vphi)
  = \frac{e\delta}{4\hbar}G_LG_R\log(\Delta/E_{g0})\sin\vphi.
 \label{I0phi}
\eeg
Here we denote current in the absence of Coulomb interaction by $I_0(\vphi)$.
Suppression of the current by the Coulomb interaction becomes stronger as the
phase on the junction increases. Qualitatively, the bare minigap $E_g(\vphi)$
decreases and the charging effects win the competition with proximity further
suppressing the current. In the symmetric junction ($G_L = G_R$) the proximity
effect vanishes completely as $\vphi$ approaches $\pi$. The weak interaction
approximation becomes invalid in this limit even if it is correct for small
$\vphi$.

\begin{figure}
\includegraphics[width=0.95\columnwidth]{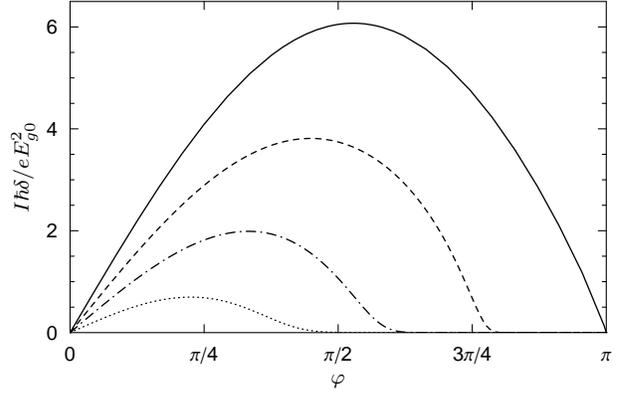}
\caption{Fig.~1. Current vs. phase for the symmetric junction. Solid curve
illustrates the case $E_C = 0$. It is given by the expression~(\ref{I0phi}).
Other curves correspond to $E_C\delta/E_{g0}^2 = 0.5,\,1.5,\,2.5$. At small
values of this parameter and at small $\vphi$ the current is given
by~(\ref{Iweak}). At larger $\vphi$ charging effects are always strong and the
current is exponentially suppressed~(\ref{Istrong}). We assume $G_L = G_R = 20$
and $\log(\Delta/E_{g0}) = 5$.}
\label{Fig_If}
\end{figure}

\emph{Strong Coulomb blockade.} For small values of the parameter $q$ we
calculate the ground state of~(\ref{hamilton}) perturbatively: $E_0 = -E_C
q^2/2(1-N^2)$. The equations~(\ref{q},\ref{selfconsist}) give
\be
 \tEg
  = 2\Delta\exp\left[-\frac{2E_C\delta}{E_g^2(\vphi)}(1 - N^2)\right],
 \quad
 q
  = \frac{2\tEg(1-N^2)}{E_g(\vphi)}.
 \label{tEg0}
\ee
Exponentially small minigap survives at $T=0$ when Coulomb blockade is strong.
Josephson current is  exponentially small as well:
\bem
 I(\vphi)
  = \frac{2e\delta^2E_C\Delta^2}{\hbar E_g^4(\vphi)}G_LG_R(1-N^2)\\
    \times\exp\left[-\frac{4E_C\delta}{E_g^2(\vphi)}(1-N^2)\right]\sin\vphi.
 \label{Istrong}
\eem

\begin{figure}
\includegraphics[width=0.95\columnwidth]{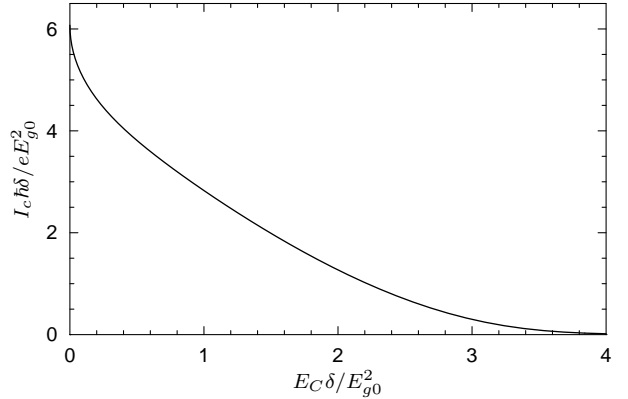}
\caption{Fig.~2. The dependence of the critical current on $E_C$ for the
symmetric junction. The parameters are $G_L = G_R = 20$, $\log(\Delta/E_{g0}) =
5$.}
\label{Fig_Icrit}
\end{figure}

We solve numerically the system of equations~(\ref{q},\ref{selfconsist}) and
plot the dependence of the current on the phase difference in
Fig.~\ref{Fig_If}. Critical current $I_c =\max_{\vphi}I(\vphi)$ as function of
charging energy is shown in Fig.~\ref{Fig_Icrit}.

\begin{figure}
\includegraphics[width=0.95\columnwidth]{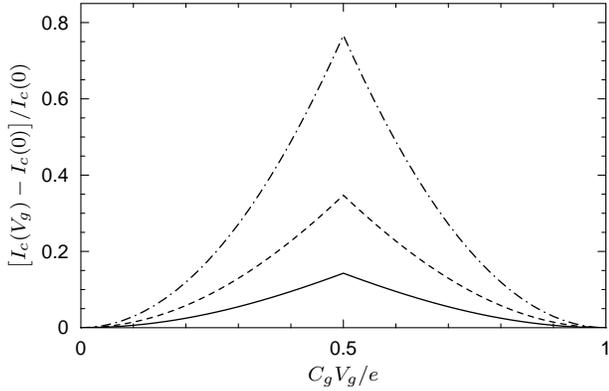}
\caption{Fig.~3. Relative increase of critical current by gate voltage for
$E_C\delta/E_{g0}^2$ equal to $1.5$, $2$, and $2.5$. This effect gets stronger
at larger charging energies. Other parameters of the junction are $G_L = G_R =
20$, $\log(\Delta/E_{g0}) = 5$.}
\label{Fig_Igate}
\end{figure}

The gate voltage enhances both the minigap and the current [see
Fig.~\ref{Fig_Igate}]. Large Coulomb energy makes the charge of the grain to be
nearly conserving quantity. Ground state corresponds to zero charge and is
separated by the gap $4E_C$ from the excited states with charge $\pm 2e$.
States with odd charge are ineffective because electrons tunnel from leads by
pairs. This situation changes when the gate voltage approaches $e/2C$. The gap
between ground state and excited state gets twice smaller assisting tunneling
of Cooper pairs. At higher gate voltage the ground state carries odd charge and
the critical current starts to diminish. The increase of current with gate
voltage is analogous to that studied in~\cite{Glazman94} where the similar
setup with the superconductive grain was considered.

\begin{figure}
\includegraphics[width=0.95\columnwidth]{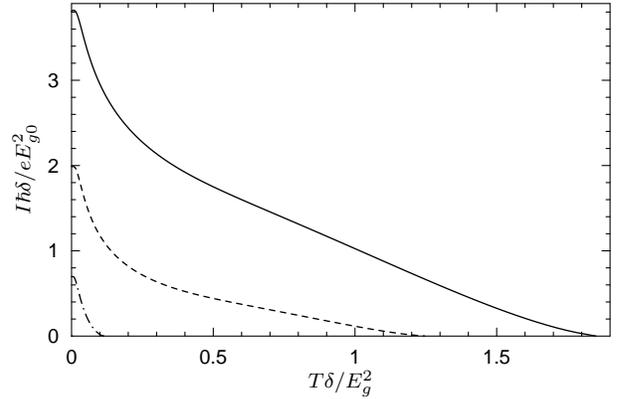}
\caption{Fig.~4. Critical current vs. temperature for $E_C\delta/E_{g0}^2 =
0.5,\,1.5,\,2.5$. Other parameters of the junction are $G_L = G_R = 20$,
$\log(\Delta/E_{g0}) = 5$.}
\label{Fig_ITemp}
\end{figure}

Now we turn to the thermodynamic properties of the junction. The temperature
dependence of the critical current is found numerically and depicted in
Fig.~\ref{Fig_ITemp}. At some temperature both the minigap and the Josephson
current disappear. As temperature approaches its critical value $T_c$ the
parameter $q$ becomes arbitrarily small. This allows to expand the free energy
of the $K$ degree of freedom: $F(q,T) = F(0,T) - E_C
\beta(N,T) q^2$. The coefficient $\beta(N,T)$ may be found with the help of
perturbation theory for the Hamiltonian~(\ref{hamilton}).
\bem
 \beta(N,T)
  = \frac{\sum_{-\infty}^\infty e^{-(n-N)^2E_C/T} /(1-(n-N)^2)}
         {2\sum_{-\infty}^\infty e^{-(n-N)^2E_C/T}}\\
  = \begin{cases}
     1/[2(1-N^2)], &T \ll E_C; \\
     E_C/T - (2/3)(E_C/T)^2, &T \gg E_C.
    \end{cases}
  \label{beta}
\eem
Note that $\beta$ is non-monotonous function of temperature. At large
temperature highly excited levels of the Hamiltonian insensitive to the
$q$-perturbation play the main role. Thus $\beta$ goes to zero in this limit.
At small temperature the phase $K$ is almost frozen at the ground state. The
$q$-term mixes two lowest excited states. As temperature grows these two states
begin to contribute to the free energy increasing its dependence on $q$. When
$N$ approaches $1/2$ the ground state becomes degenerate and $\beta$ falls
monotonously with temperature.

The expression~(\ref{q}) in the limit $\tEg \ll T$ gives
\be
 q
  = \frac{E_g\tEg}{E_C\delta}\log\frac{2\gamma\Delta}{\pi T}.
 \label{qT}
\ee
Here $\log\gamma \approx 0.577$ is the Euler constant. The self-consistency
condition~(\ref{selfconsist}) in the limit of small $q$ takes the form $\tEg =
E_g(\vphi)\beta q$. Substituting this equation into Eq.~(\ref{qT})  we find  for the
critical temperature:
\be
 T_c
  = \frac{2\gamma}{\pi}\Delta
    \exp\left[-\frac{E_C\delta}{E_g^2(\vphi)\beta(N,T_c)}\right].
 \label{Tc}
\ee
The same equation may be obtained by the expansion of~(\ref{freeen}) in powers
of $\tEg$ and setting the coefficient of $\tEg^2$ to zero. It can be checked
that the forth-order term of this expansion always remains positive. This
justifies our assumption that $\tEg$ vanishes continuously at the critical
temperature.

In the regime of strong Coulomb interaction critical temperature appears to be
much less than $E_C$. Taking low temperature asymptotic of $\beta(N,T)$ we find
that $T_c = (\gamma/\pi)\tEg(T=0)$, where $\tEg(T=0)$ is given by
Eq.~(\ref{tEg0}). This is the BCS relation between the gap and the critical
temperature. The phase of the grain strongly fluctuates and is mainly
independent of the phase in the leads. The only effect of superconductive leads
is a weak effective attraction in the Cooper channel which leads to formation
of very weak BCS-like state.

In the opposite limit of weak Coulomb interaction we employ high temperature
asymptotic of $\beta$ and find
\be
 T_c
  = \frac{E_g^2(\vphi)}{\delta}
    \log\frac{2\gamma\Delta\delta}{\pi E_g^2(\vphi)} - \frac 23 E_C.
\ee

\begin{figure}
\includegraphics[width=0.95\columnwidth]{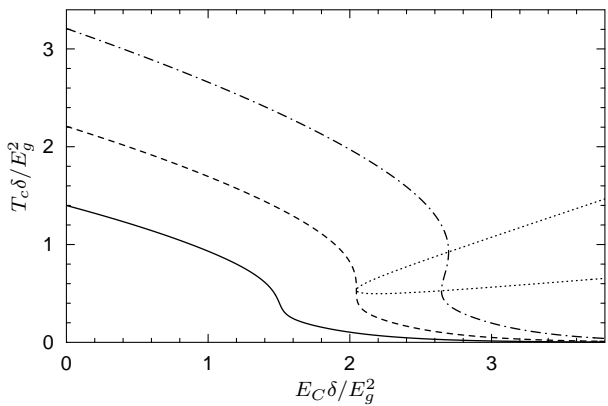}
\caption{Fig.~5. The critical line demonstrating the dependence of critical
temperature on $E_C$. The three curves are plotted for $\Delta\delta/E_g^2 =
5,\,17.7,\,70$. At large enough values of $\Delta$ this dependence may not be
single-valued. Dotted line shows the turning points of critical temperature.
Small disk denotes the critical point where the dotted line touches the $T_c$
vs. $E_C$ curve. This critical point gives the following critical values:
$T_c^*\delta/E_g^2 = 0.54$, $E_C^*\delta/E_g^2 = 2.05$, $\Delta^*\delta/E_g^2 =
17.7$.}
\label{Fig_Tc}
\end{figure}

The whole dependence of the critical temperature on parameters is shown in
Fig.~\ref{Fig_Tc}. It is possible that the equation~(\ref{Tc}) has more than
one solution at a given value of $E_C$. This implies  reentrant behavior of the
minigap and critical current as functions of temperature. Mathematically, it is
due to nonmonotonic behavior of the function $\beta(T)$. Physical explanation
is most simple in charge rather than phase representation of the
Hamiltonian~(\ref{hamilton}). Two excited states with charge $\pm e$ have equal
charging energies. Tunneling of Cooper pair, that switches between these two
states, is not blocked by Coulomb interaction. Finite temperature may excite
the system to one of this states leading to the temperature-assisted proximity
effect and the enhancement of the minigap. The Hamiltonian ~(\ref{hamilton})
conserves the parity of electron number. Thus thermalization of the system
implies some parity-breaking  processes (e.g. single electron tunneling with
energy above $\Delta$), that may take a long time.

\begin{figure}
\includegraphics[width=0.95\columnwidth]{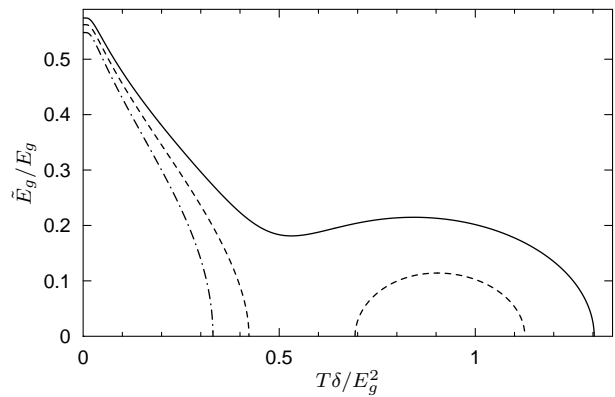}
\caption{Fig.~6. The minigap in the normal dot as a function of temperature.
Solid line is plotted for $E_C\delta/E_g^2 = 2.6$. Dashed line illustrates the
case $E_C\delta/E_g^2 = 2.67$. Chain line is for $E_C\delta/E_g^2 = 2.75$. All
curves are plotted for  $\Delta = 70 E_g^2/\delta$. The critical temperature as
a function of $E_C$ for the same value of $\Delta$ is shown by the chain line
in Fig.~\ref{Fig_Tc}.}
\label{Fig_Eg}
\end{figure}

At large enough values of  $\Delta > \Delta^* = 17.7E_g^2/\delta$, reentrant
behavior  of the minigap with temperature was found in some (dependent on
$\Delta/\Delta^*$ ratio) interval of the Coulomb parameter $E_C\delta/E_g^2$,
cf. Fig.~\ref{Fig_Eg}. Fine tuning of the parameter $E_C\delta/E_g^2$ can be
achieved by an appropriate phase bias, cf. Eq.~(\ref{Egphi}).

To conclude, we have described the Josephson effect in a nanoscale SINIS
junction modified by the Coulomb interaction. The most important feature of
SINIS structure (in comparison with usual SIS structure) is that it can
demonstrate both good metallic conductance in the normal state and Coulomb
blockade of Josephson current at very low temperatures, since both the
conditions $G_{L,R} \gg 1$ and $E_J = \frac{\hbar}{2e}I_c \leq E_C$ can be
fulfilled simultaneously. We calculated the current-phase characteristic of the
junction in both weak and strong Coulomb blockade limit. The enhancement of the
current by the gate voltage is predicted. The temperature dependence of the
critical current and of the  minigap induced in the normal part of the junction
was found. A grain of noble metal with size about 50 nm connected to Nb
superconductive electrodes by tunnel oxide barriers with transparency per
channel of the order of $10^{-5}$ could present an example of the studied
system with $E_J \sim E_C \sim 1 K$. We are grateful to T.\,Kontos and
Ya.\,Fominov for useful discussions. This research was supported by the Program
``Quantum Macrophysics" of the Russian Academy of Sciences, Russian Ministry of
Science and RFBR under grant No.\ 04-02-16348. P.\,M.\,O. acknowledges
financial support from the Dynasty Foundation and the Landau Scholarship (KFA
Juelich).

\end{document}